\documentstyle[12pt]{article}

\title{Quantum theory for the Standard Model}

 \author{M. Talon
 \thanks{L.P.T.H.E. Universit\'e Paris VI (CNRS UA 280),
 Bo\^{\i}te 126, Tour 16, $1^{er}$ \'etage,
 4 place Jussieu, F-75252 PARIS CEDEX 05} }

\date{September 1993}

\begin{document}

\begin{titlepage}
\renewcommand{\thepage}{}
\maketitle
\vskip 2cm

\begin{abstract}
These lectures present some basic facts in field theory necessary to
understand the quantum theory of the Standard Model of weak
and electromagnetic interactions.
\end{abstract}
\vfill
%\centerline{\epsffile{logo.eps}}
Saint Fran\c cois Guadeloupe \hfill\break
PAR LPTHE 93-58 \hfill
Work sponsored by CNRS: URA 280

\end{titlepage}
\renewcommand{\thepage}{\arabic{page}}

\tableofcontents
\cleardoublepage

\section{Introduction}

The aim of these lectures is to introduce the reader to the
theoretical setup of the Standard Model of weak and electromagnetic
interactions. This entails a presentation of the perturbative
expansion of quantum field theory, which is also a prerequisite to
the lectures of R. Stora on renormalization of gauge field theory.
Finally this standard view on the Standard Model is an introduction
to the lectures of D. Kastler which will present a modern viewpoint
on this theory based on the ideas on non-commutative geometry
developed by A. Connes.

The quickest way (hence the clearest) to the perturbative expansion is
the quantification by path integration, historically introduced by
R. Feynman in this context. This approach leads to some problems with
respect to a rigorous definition of the path integral. These problems
have been overcome, notably by an analytic continuation to the
Euclidean region; in turn this allows powerful developments in
Constructive Quantum Field Theory, that we shall not be concerned about
. More pragmatically path integration does the magical job of
dissolving the ordering problems of quantum mechanics; we show that
this is of course an illusion and that these problems are simply
transformed into discretization ones.

Another important issue is the question of causality in Field Theory.
It is frequently unclear how the correct $i\epsilon$ prescription
appears in the propagators, and why the path integral of product of
operators yields a T--product. We have chosen to show, following the
original work of R. Feynman~\cite{F1,F2}, that a sufficiently thorough study of
a
simple example, the forced harmonic oscillator, gives the correct
answers.

We then introduce the important generating functionals of connected
diagrams and one--particle--irreducible diagrams. They are
combinatorial objects of much use in the discussion of the symmetries
of the considered theories and their renormalization. Since their
properties are relegated to exercises in the usual textbooks we
present proofs of the main facts. These combinatorial objects are also
useful in the applications of field theory to Statistical
Mechanics, see for example~\cite{F,ID}.

Finally we briefly discuss an essential ingredient of the Standard
Model, i.e., the Higgs mechanism, but we do not offer a discussion of the
quantification of gauge theories~\cite{Lee} since this would duplicate R.
Stora's
lectures. Lack of space prevents us to introduce the subject of
anomalies, that is breaking of gauge invariance by quantum effects due
to the divergences of quantum field theory, which lead to an
important constraint on the construction of the Standard Model, namely
that the families must be complete.

All these subjects are treated fully in the standard textbooks on
Quantum Field Theory, either the classical ones~\cite{BD,BS}, or the modern
ones~\cite{IZ,Z,CL} which moreover offer a complete discussion of
renormalization group theory, which is essential to understand
QCD, i.e., the present theory of strong interactions. Streamlined accounts can
be found in~\cite{KH,R} and notably in the famous report of E. Abers and
B. Lee~\cite{AL}. Finally we must mention the beautiful lectures of S.
Coleman~\cite{C} which cover roughly the same subjects as presented below.

\section{The path integral}
\setcounter{equation}{0}

\subsection{Path integral representation of the evolution operator}

We work in the Heisenberg representation, for a one dimensional system,
and denote  $Q(t)$  the position operator at time  $t$.
$$ Q(t)=e^{iHt} Q e^{-iHt} $$
Let us recall that the quantum states have no time evolution in this
representation. We denote $|q,t>$    the fixed eigenstate of  $Q(t)$  for the
eigenvalue $q$,
where $t$ is given some fixed value. If $|q>$  is such that
$Q|q>=q|q>$, then $|q,t>=e^{iHt}|q>$.
Finally, for any state $|\psi >$, $<q,t|\psi >=<q|e^{-iHt}|\psi>$
is the Schr\"{o}dinger representation of this state at the time $t$.

The evolution operator is the amplitude to go from  to $|q,t>$ to $|q',t'>$,
i.e., knowing that the particle is sitting at $q$  at time $t$,
the amplitude to find it at $q'$ at time $t'$.
$$<q',t'|q,t>=<q'|e^{-iH(t'-t)}|q>$$
For any decomposition of the time interval $(t,t')$ we have:
$$t=t_0<t_1<\cdots<t_n<t_{n+1}=t'$$
$$<q',t'|q,t>=\int dq_1\cdots dq_n <q',t'|q_n,t_n><q_n,t_n|\cdots|q,t>$$
since for any $t_j$  the states $|q_j,t_j>$  are a complete set.

For  $\epsilon$  small one traditionally performs the following approximations,
hardly justified in general:
$$<q',t+\epsilon|q,t>=<q'|e^{-i\epsilon H(P,Q)}|q>\simeq <q'|1-i\epsilon
H(P,Q)|q>$$
$$= \int {dp \over 2\pi}e^{ip(q'-q)}\left( 1-i\epsilon H(p,{q+q' \over
2})\right)$$
(under appropriate ordering of the operators $P$  and  $Q$ in   $H$, as
explained later
on)
$$\simeq \int {dp \over 2\pi}e^{ip(q'-q) -i \epsilon H(p,{q+q' \over 2})}$$
Notice that all operators have disappeared in the final expression. This
approximation has
first been noted by Dirac.

Finally one gets with  $\epsilon={t'-t\over n+1}$   the expression of the
evolution
operator $<q',t'|q,t>$ in the form:
$$\int dq_1\cdots dq_n {dp_0 \over 2 \pi}\cdots {dp_n \over 2 \pi}
\exp\,i\epsilon  \sum_{k=0}^n \left\{ p_k {q_{k+1}-q_k \over \epsilon}
-H(p_k,{q_k+q_{k+1} \over 2}) \right\}$$ Of course, one defines:
$$\dot{q}_k={q_{k+1}-q_k \over \epsilon},\quad \overline{q}_k={q_k+q_{k+1}
\over 2}$$
so that in the limit $n \to \infty$  one gets the path integral expression of
the evolution
operator:
\begin{equation}
<q',t'|q,t>=\int \prod_\tau\, { dp(\tau)\, dq(\tau) \over 2\pi}
e^{i \int_t^{t'} (p\dot{q} -H(p,\overline{q})) d\tau}\label{path1}
\end{equation}
Notice that in this expression   $H$  may well be explicitly time dependent.
Also notice that this path integral takes the form of an integral over
partially constrained paths in phase space of the exponential of the classical
action, averaged with the invariant measure of phase space. So it seems naively
well behaved under canonical transformations when ordering and discretization
problems are not taken into account.

In the particular case in which the phase space is polarized by a globally
defined choice of a configuration variable $Q$ and of the momentum $P$ and
the Hamiltonian is of the form:
$$H(P,Q)={1 \over 2}P^2+V(Q)$$
notice that there are no ordering problems in the above approximation
steps, and moreover one can explicitly integrate over all variables,
using the Gaussian integration formula:
$$\int {dp\over 2\pi} e^{i\epsilon[p\dot{q}-{p^2\over 2}]}=
{1\over \sqrt{2\pi i\epsilon}}e^{i\epsilon{\dot{q}^2\over 2}}$$
One ends up with:
$$<q',t'|q,t>=\int dq_1\cdots dq_n {1\over(2\pi i \epsilon)^{n+1\over 2}}
e^{i\epsilon \sum_{k=0}^n({1\over2}\dot{q}_k^2-V(q_k))}$$
In this situation one introduces the Lagrangian
$$L(q,\dot{q})={1\over2}\dot{q}^2-V(q)$$
so that we get in the limit $n\to\infty$:
\begin{equation}
<q',t'|q,t>=\int \prod_\tau {dq(\tau)\over\sqrt{2\pi i\epsilon}}
e^{i\int_t^{t'} L(q,\dot{q}) d\tau} \label{path2}
\end{equation}
This is the form of the Path Integral first written by Feynman. Notice that
$S=  \int_t^{t'} L(q,\dot{q}) d\tau$
is the classical action along the path $q(\tau)$ in configuration space leading
from $q$ at $t$ to $q'$ at $t'$ and one averages  over all such paths.

This Feynman formula can be proven rigorously assuming $H$ to be
self-adjoint, by an application of the Trotter product formula, as first
shown by Nelson~\cite{N}.
The above more ``general'' one~(\ref{path1}) has no such solid
foundation. In turn the Feynman formula can be used to prove rigorous
results. In this context one generally works in the Euclidean
formulation in which time is rotated to pure imaginary values. Then
$$\prod_\tau {dq(\tau)\over\sqrt{2\pi \epsilon}}
e^{-\int_t^{t'} \dot{q}^2/2\, d\tau}$$
can be shown to be a well--defined measure on the space of paths,
first introduced by N. Wiener, with which one averages the
potential--dependent functional of the path.
An extended account of these applications can be found in the
book of B. Simon~\cite{Si}. At a more ``physical'' level Feynman has
discovered a nice application of the path integral in his famous study
of the polaron problem, discussed as well as other interesting
applications to Statistical Mechanics in his book~\cite{F}.

\subsection{The ordering problem}

As we have seen before, the operators $P$ and $Q$ present in the evolution
operator disappear in the computation of the path integral and one ends up
with purely numerical quantities. The crucial steps in the computation is
reached when we write:
$$<q'|H(P,Q)|q>=\int {dp\over 2\pi}e^{ip(q'-q)}H(p,{q+q'\over 2})$$
under ``appropriate'' ordering hypothesis.

Notice that if one writes $H(P,Q)$  with all the $P$'s  at the left and all the
$Q$'s   at the right, as in Faddeev~\cite{Fa}, then trivially:
$$<q'|P^nQ^m|q>=\int dp <q'|P^n|p><p|Q^m|q>$$
$$=\int {dp\over 2\pi}e^{ip(q'-q)} p^nq^m$$
so that one ends up with $H(p,q)$   instead of $H(p,{q+q'\over 2})$.
The problem is that $H(P,Q)$  is then not
obviously self-adjoint. We see that ordering ambiguities translate into
discretization problems in the Path Integral formulation.

Let us now define the so--called Weyl ordering which leads to the above
``mid--point splitting''. We shall only concern ourselves with functions of
the form $f(q)p^r$   with $r=0,1,2$ the only ones of physical interest.
A more general discussion can be found in the book of Berezin~\cite{B}.
For $r=0$ there is no problem. Then we define following T.D. Lee~\cite{L}:
\begin{eqnarray}
\relax [Q^nP]_W &=& {1\over n+1}[Q^nP+Q^{n-1}PQ+\cdots +PQ^n] \nonumber\\
\relax [Q^nP^2]_W &=& {2\over
(n+1)(n+2)}\sum_{l,m}Q^{n-l-m}PQ^lPQ^m\label{Weyl1}
\end{eqnarray}
i.e., one averages over all possible orders.

Then by using the commutation relations:
$$PQ^l=-ilQ^{l-1}+Q^lP$$
one can show by brute force computation that for any polynomial $f$:
\begin{eqnarray}
\relax [f(Q)P]_W &=& {1\over 2}[f(Q)P+Pf(Q)] \nonumber\\
\relax [f(Q)P^2]_W &=& {1\over 4}[f(Q)P^2+2Pf(Q)P+P^2f(Q)]\nonumber\\
\relax [Q^nP^r]_W &=& {1\over 2^n}\sum_l C_n^l\, Q^{n-l}P^rQ^l \label{Weyl2}
\end{eqnarray}

As a matter of fact, each of the expressions in~(\ref{Weyl1},\ref{Weyl2})
for $r=1$ can be reduced to  $Q^nP-in/2\,Q^{n-1}P$,
and each of the expressions for $r=2$ boils down to
$Q^nP^2-inQ^{n-1}P-1/4\,n(n-1)Q^{n-2}$,
but notice that Weyl ordered expressions are obviously self-adjoint.

Now we can easily compute for $r=0,1,2$ the required kernel:
\begin{eqnarray*}
\lefteqn{<q'|[Q^mP^r]_W|q>=}\\
&=& {1\over 2^m}\,\sum_l\int {dp\over 2\pi} C_m^l e^{ip(q'-q)}q'^{m-l}p^rq^l\\
&=& \int {dp\over 2\pi} e^{ip(q'-q)}p^r({q+q'\over 2})^m
\end{eqnarray*}
yielding the mid--point splitting as promised earlier.

\subsection{The time ordered products}

One can give path integral formulations of more general expectation
values such as:
$$<q',t'|F_n(t_n)\cdots F_1(t_1)|q,t>$$
where $F_j(t_j)$ are Heisenberg operators, i.e., functions of $P(t_j)$  and
$Q(t_j)$ assumed for example Weyl ordered. We also assume
$$t<t_1<\cdots t_n<t'$$

Then one can choose time splittings of the interval $(t,t')$ such that all the
$t_j$'s occur in the decomposition.
Proceeding as above we shall express the above expectation value in the form:
$$\int \prod_l dq_l <q',t'|\cdots <q_{k+1},t_k|F_j|q_k,t_k>
<q_k,t_k|q_{k-1},t_{k-1}>\cdots |q,t>$$
where $t_k$ is the same time occurring in $F_j$. Since $F_j$ is Weyl ordered
one has:
$$<q_{k+1},t_k|F_j|q_k,t_k>=\int {dp_k\over 2\pi} e^{ip_k(q_{k+1}-q_k)}
F_j(p_k,{q_k+q_{k+1}\over 2})$$
and in the limit we get:
$$\int\prod_\tau  { dp(\tau)\, dq(\tau) \over 2\pi}
\prod_j F_j(p(t_j),\overline{q}(t_j))\,
e^{i \int_t^{t'} (p\dot{q} -H(p,\overline{q})) d\tau}$$
Notice that for the given path $(p(\tau),q(\tau))$ one has to evaluate the
function
$F_j$ at the position corresponding to the time $t_j$.

In the case in which $H=1/2\,P^2+V(Q)$  and   $F_j=[A_j(Q)+B_j(Q)P]_W$
one can easily eliminate the integrations over
variables $p_k$  and obtain a Feynman type formula for the expectation value:
$$\int \prod_\tau {dq(\tau)\over\sqrt{2\pi i\epsilon}}
\prod_j F_j(\dot{q}(t_j),\overline{q}(t_j)) \,e^{iS}$$
Of course the product $\prod_j F_j$  is now a product of numbers, i.e., does
not depend on
the order of the times $t_j$.  Hence the preceding derivation shows that the
above path integral is in general the expression of
$$<q',t'|F_{\sigma(n)}\cdots F_{\sigma(1)}|q,t>$$
where the permutation $\sigma$ is such that:
$$t<t_{\sigma(1)}<\cdots t_{\sigma(n)}<t'$$
Such a permuted product is called a time-ordered product, and we have
shown that:
\begin{equation}
<q',t'|T\left( F_n(t_n)\cdots F_1(t_1)\right)|q,t>=
\int {\cal D}q \, F_1 \cdots F_n \, e^{iS} \label{tprod}
\end{equation}
where ${\cal D}q$ denotes the appropriate product of the $dq(\tau)$.

Remark that the appearance of the T--product in this expression ultimately
comes from the assumption that the dissection of the time interval $(t,t')$  is
time--ordered, which can be taken as a definition of the Path Integral,
while this is compulsory in the Euclidean formulation. Moreover the
appearance of the mid--point splitting in this
formula leads to a more refined version of the ordinary T--product,
called the  T$^*$--product. The subtlety lies in the definition of T--product
for
coincident points. The implications are developed for example in Adler's
lectures~\cite{A}.

\subsection{The forced harmonic oscillator}

We shall now follow Feynman and indicate how one can compute the path
integral for a forced harmonic oscillator, a situation that directly
generalizes to Field Theory. Let us take:
$$L={1\over 2}(\dot{q}^2-\omega^2q^2)+J(t)q$$
where $J(t)$ is some external, time--dependent source (exemplifying a
time-dependent Hamiltonian).

We want to compute   $\int {\cal D}q e^{iS}$. When $\omega=0$, the
free--particle situation, the answer is immediate using Fourier expansion.
For example for $J=0$ one can write:
$$<q',t'|e^{-iH(t'-t)}|q,t>=\int {dk\over 2\pi}
e^{ik(q'-q)-{i\over 2}k^2(t'-t)}$$
$$= {1\over\sqrt{2i\pi T}}e^{i(q'-q)^2\over 2T}\quad{\rm where}\quad T=t'-t$$

When $\omega \neq 0$ the standard method is to expand around the classical
solution. So let us write $q=q_0+y$ with $q_0(t)=q$, $q_0(t')=q'$,
$y(t)=y(t')=0$,
where $q_0$ extremizes the action, i.e., $S(q_0+y)$  has no linear term in $y$.
Here the
expansion is exact:
$$S(q)=S(q_0)+\int_t^{t'}(\dot{q}_0\dot{y}-\omega^2q_0y+Jy)d\tau
+\int_t^{t'} {1\over 2}(\dot{y}^2-\omega^2y^2)d\tau$$
Since $y$ vanishes at end points one can write
$\int_t^{t'}\dot{q}_0\dot{y}d\tau=-\int_t^{t'} y \ddot{q}_0 d\tau$ so that
$q_0$
obeys the classical equation of motion:
$$\ddot{q}_0+\omega^2 q_0=J$$

In order to find the classical solution it is convenient to use the method of
Green's functions, i.e., to first consider the case in which $q=q'=0$  and
$J(\tau)=\delta(\tau-\sigma)$.
Then, setting $T=t'-t$ , the solution, an appropriate combination of sinusoids,
is:
$$G(\tau,\sigma)=-{1\over\omega\sin\omega T}\left\{
\begin{array}{l}
\sin\omega(t'-\sigma)\;\sin\omega(\tau -t)\quad \tau<\sigma\\
\sin\omega(\sigma-t)\;\sin\omega(t'-\tau)\quad\tau>\sigma
\end{array}
\right.$$
For general $J$ the solution is obviously $\int_t^{t'}
G(\tau,\sigma)J(\sigma)d\sigma$
and finally one fulfills the correct boundary conditions by adding a pure
sinusoid. One gets:
$$\begin{array}{ll}
q_0(\tau)=&
-{1\over\omega\sin\omega T}
\begin{array}{l}
\Bigl\{\sin\omega(t'-\tau)\int_t^\tau \sin\omega(\sigma-\tau)J(\sigma)d\sigma\\
 \; +\sin\omega(\tau
-t)\int_\tau^{t'}\sin\omega(t'-\sigma)J(\sigma)d\sigma\Bigr\}
\end{array}\\  &
+{q\over\sin\omega T}\sin\omega(t'-\tau)+{q'\over\sin\omega
T}\sin\omega(\tau-t)
\end{array}$$
Then the path integral reads:
$$<q',t'|q,t>=e^{iS_J(q_0)}\,\int{\cal D}y\,e^{iS_0(y)}$$
The classical action is computed according to:
$$S_J(q_0)=\int_t^{t'}[{1\over 2}(\dot{q}_0^2-\omega^2q_0^2)+Jq_0]d\tau$$
$$={1\over 2}[q_0\dot{q}_0]_t^{t'}+{1\over 2}\int_t^{t'}Jq_0d\tau$$
by integrating by parts  (the boundary terms do not cancel).
Substituting the above expression of $q_0$ one notices that the two terms have
the
same value so the $1/2$ disappears.

It remains to compute the path integral for the free harmonic oscillator.
This can be done by following the definition as a limiting procedure, see
Schulman~\cite{Sc}, but we shall content ourselves with a quick computation.
Setting $t=0$ $t'=T$ and, due to the boundary conditions:
$$y(\tau)=\sum_{n=1}^\infty y_n\sin{n\pi\tau\over T}$$
so that
$$S_0(y)={T\over 4}\sum_{n=1}^\infty
({\pi^2n^2\over T^2}-\omega^2)y_n^2$$
one writes ${\cal D}y=\prod_n dy_n$ up to some normalizing factor,
so that the path integral
reduces to independent Gaussian integrations leading to:
$$\prod_{n=1}^\infty {1\over\sqrt{1-{\omega^2T^2\over\pi^2n^2}}} \quad{\rm or}
\quad\sqrt{{\omega T \over\sin\omega T}}$$ up to some constants.
One adjusts the constant by comparing with
the case $\omega=0$ in which the computation is trivial, and one finally gets:
\begin{equation}
<q',t'|q,t>=\sqrt{\omega\over 2\pi i \sin\omega T}\;e^{i\Omega}\label{harm1}
\end{equation}
\begin{eqnarray}
\Omega &=& {\omega\over 2\sin\omega T}[(q^2+{q'}^2)\cos\omega T
-2qq']\nonumber\\
&&+{q'\over\sin\omega T}\int_t^{t'} J(\tau)\sin\omega(\tau-t)d\tau\nonumber\\
&&+{q\over\sin\omega T}\int_t^{t'} J(\tau)\sin\omega(t'-\tau)d\tau \\
&&-{1\over\omega\sin\omega T}\int_t^{t'}d\sigma
\int_t^\sigma d\tau J(\sigma)J(\tau)\sin\omega(t'-\tau)\sin\omega(\sigma-t)
\nonumber\label{harm2}\end{eqnarray}

It is interesting to consider the limiting form of this result when $\omega\to
0$,
and to compare with the direct computation for $\omega =0$.

\subsection{Vacuum expectation values}

We shall introduce now the objects of interest in the generalization to
Quantum Field Theory, i.e., vacuum expectation values. In general, let us
consider a time--independent system coupled to a source $J(\tau)$ such that
$J(\tau)\neq 0$ only for $t<\tau<t'$ and finally take $T\ll t$  and $T'\gg t'$.
Then:
$$<Q',T'|Q,T>^J=\int dqdq'<Q',T'|q',t'><q',t'|q,t>^J<q,t|Q,T>$$
But outside of $(t,t')$    we have free motion under the Hamiltonian $H$ with
spectrum  $(E_n,\phi_n)$. Hence:
$$<q,t|Q,T>=<q|e^{-iH_0(t-T)}|Q>$$
$$=\sum_n \phi_n(q) \phi_n^*(Q) e^{-iE_n(t-T)}$$
Let us now formally assume some analytic continuation into Euclidean
space, i.e.:
$$T\to -(1-i\epsilon)\infty \quad{\rm and}\quad T'\to +(1-i\epsilon)\infty$$
This has the effect of projecting on the ground state:
$$<q,t|Q,T>\simeq \phi_0(q)\phi_0^*(Q)e^{-iE_0(t-T)}$$
We see that:
\begin{eqnarray}
W[J] &\equiv& {<Q',T'|Q,T>^J \over e^{-iE_0(T'-T)}\phi^*_0(Q)\phi_0(Q')}
\nonumber \\
&=& \int dqdq' \phi_0(q,t)\phi_0^*(q',t')<q',t'|q,t>^J \label{wdej}
\end{eqnarray}
in which we set $\phi_0(q,t)=e^{-iE_0t}\phi_0(q)=<q,t|\phi_0>$
i.e., the Schr\"{o}dinger wave-function at time $t$ corresponding to the ground
state $|\phi_0>$. In particular this shows that this vacuum expectation value
does
not depend on $t$ and   $t'$.

We apply this idea to the forced harmonic oscillator, where:
$$\phi_0(q)=\left({\omega\over\pi}\right)^{1\over 4}e^{-\omega q^2/2}$$
We need to compute:
$$W[J]={\omega\over\pi}{e^{i\omega T/2}\over\sqrt
{2i\sin\omega T}}\int dqdq'e^{-\omega(q^2+{q'}^2)/2+i\Omega}$$
where $\Omega$  is the previously defined expression. This is a Gaussian
integration which separates when setting $q=u+v$, $q'=u-v$. After a lengthy
computation,
one observes that all factors like $\sin \omega T$  (here $T=t'-t$) cancel,
and one gets the remarkably simple result:
$$W[J]=\exp\{-i\int_t^{t'}d\sigma \int_t^\sigma d\tau
J(\sigma){e^{-i\omega(\sigma -\tau)}\over 2i\omega}J(\tau)\}$$

The point of this computation, due to Feynman is that one ends up with the
correct Feynman propagator between $J(\sigma)$  and $J(\tau)$  as a result of
having correctly taken into account the boundary conditions, and without any
appeal to
continuations into the Euclidean region. This is to be compared to the analysis
in Ramond's book~\cite{R}, in which a convergence factor is used.

Due to the vanishing of $J$ outside $(t,t')$ one can write:
\begin{eqnarray}
W[J] &=& \exp\{ {-i\over 2}\int_\infty^\infty \int_\infty^\infty
d\sigma d\tau J(\sigma) D_F(\sigma - \tau) J(\tau)\} \nonumber\\
D_F(\sigma) &=& {1\over 2i\omega} [\theta(\sigma)e^{-i\omega\sigma}
+\theta(-\sigma)e^{i\omega\sigma}]\label{propag}\\
&=& \int_\infty^\infty {dE\over 2\pi}{e^{i\sigma E}\over E^2-\omega^2
+i\epsilon}
\nonumber\end{eqnarray}
This means that the harmonic oscillator induces an effective interaction
between $J(\sigma)$ and $J(\tau)$ such that positive frequencies are propagated
forward in
time, and negative ones backwards.

The same result is obtained straightforwardly by setting
$\omega^2\to\omega^2-i\epsilon$
in the Lagrangian, i.e., introducing a convergence factor in the Path Integral,
which in effect destroys the time-reversal symmetry of Quantum Mechanics.

\subsection{Generalization to quantum field theory}

The preceding discussion immediately generalizes to the case of a free
field $\phi(\vec{x},t)$  coupled to external sources.
As a matter of fact, such a field (with
mass term $\mu$ ) can be considered under Fourier transformation on $\vec{x}$,
as a
collection of harmonic oscillators $\phi_{\vec{k}}(t)$  of frequency
$\omega_{\vec{k}}=\sqrt{\vec{k}^2+\mu^2}$. The corresponding
Lagrangian density is:
$$L_J={1\over 2}(\partial_\mu \phi)^2-{\mu^2\over 2}\phi^2+J\phi$$
Without further ado we write the vacuum expectation value as:
$$W[J]=\exp\,\{ {-i\over 2}\int d^4xd^4y J(x)\Delta_F(x-y)J(y)\}$$
with the Feynman propagator:
\begin{equation}
\Delta_F(x)=\int{d^4k\over (2\pi)^4}{e^{ik.x}\over k^2-\mu^2+i\epsilon}
\label{feyprop}\end{equation}

Let us remark that time--ordered products of fields are simply obtained by
functional differentiation of  $W[J]$:
\begin{eqnarray}
<0|T(\phi(x_1)\cdots  \phi(x_n))|0>_J &=& \int {\cal D}\phi \,
 \phi(x_1)\cdots  \phi(x_n)\, e^{i\int d^4x L_J} \nonumber \\
&=& \left( {1\over i}{\delta\over\delta J(x_1)} \right)\cdots
\left( {1\over i}{\delta\over\delta J(x_n)} \right)W[J]\nonumber\\
\label{green}
\end{eqnarray}
Since $W[J]$  is known we can compute this expression, and then go to the limit
$J\to 0$. Obviously to get something non zero in this limit we have to consider
terms like:
$$\left( {1\over i}{\delta\over\delta J(x_1)} \right)e^{-{i\over 2}\int
J\Delta_F J}
= -\int dy \Delta_F(x_1-y)J(y)e^{-{i\over 2}\int J\Delta_F J}$$
$$\left( {1\over i}{\delta\over\delta J(x_1)} \right)
\left( {1\over i}{\delta\over\delta J(x_2)} \right)
e^{-{i\over 2}\int J\Delta_F J}=i\Delta_F(x_1-x_2)$$
plus terms vanishing in the limit $J\to 0$.

This means that the result is obtained by connecting all pair of points   by
propagators   in all possible ways (a propagator is a line
connecting $x_k$ to $x_l$, to which is associated the value
$i\Delta_F(x_k-x_l)$), and adding all such expressions. One
such product is called a Feynman graph, of tree type.
We have expressed the Green function:
\begin{equation}
G(x_1,\cdots,x_n)=<0|T(\phi(x_1)\cdots  \phi(x_n))|0> \label{greendef}
\end{equation}
as functional derivative of $W[J]$ when $J\to 0$. This means that
$W[J]$ is the generating function of Green's functions:
\begin{equation}
W[J]=\sum_{n=0}^\infty {1\over n!} \int \prod dx_i\,
G(x_1,\cdots,x_n)\, J(x_1)\cdots J(x_n)
\end{equation}

In the above trivial (free) situation, we have also shown that one can
write:
\begin{eqnarray*}
W[J] &=& \exp (iZ[J])\\
Z[J] &=& -{1\over 2}\int d^4x d^4y J(x) \Delta_F(x-y) J(y)
\end{eqnarray*}
We shall elaborate in the following on these important generating
functions of Field Theory.

\subsection{The perturbation theory}

When self--interaction terms are present in the Lagrangian it is impossible
to express $W[J]$ in closed form as above, and Green's functions are computed
as perturbation series in the coupling constants.

One writes  $L=L_0+L_I$ where  all terms quadratic or linear in the fields are
collected in   $L_0$ and all terms cubic and higher in $L_I$. Finally one
expands in
$\int {\cal D}\phi\,\phi(x_1)\cdots\phi(x_n)\, e^{i\int d^4x L}$
the exponential in the form:
$$e^{i\int d^4x L}=e^{i\int d^4x L_0}\,\sum_{n=0}^\infty
{1\over n!}\left(i\int d^4x L_I\right)^n$$
This brings us back to the computation, order by order, of Green's functions
for the free field.

Taking for example $L=L_0+{\lambda\over 4!}\phi^4$ where $L_0$ is
the previously defined quadratic Lagrangian density we see that
\begin{equation}
G(x_1,\cdots,x_n)=\int {\cal D}\phi\,\phi(x_1)\cdots\phi(x_n)\,
\sum_{m=0}^\infty {1\over m!}\left( {i\lambda\over 4!}
\int \phi^4(y) dy\right)^m\,e^{iS_0}
\label{pert}\end{equation}
At each order $m$ in the coupling constant we have to compute the Green
function corresponding to the product of fields
$${1\over m!} \left({i\lambda\over 4!}\right)^m\int dy_1\cdots dy_m\,
\phi^4(y_1)\cdots\phi^4(y_m)\phi(x_1)\cdots\phi(x_n)$$

Using the method of generating functions each $\phi^4(y_k)$ is converted to
$\left[{1\over i}{\delta\over \delta J(y_k)}\right]^4$  which,
encountering a product of four $J$'s in the expansion of $\exp(iZ[J])$  gives
$$\prod_{l=1}^4 J(z_l)\to 4!\prod_{l=1}^4\delta(z_l-y_k)$$
Notice that the $i$'s  cancel correctly as explained above, leaving
$i\Delta_F$ for each propagator.

In other words we get a Feynman diagram with $n$  terminals
$x_1,\cdots,x_n$  and $m$ vertices  $y_1,\cdots,y_n$
(which is then integrated other the $y_k$'s). To each vertex  $y_k$
are attached four propagators. Each propagator has value $i\Delta_F(z_i-z_j)$
and ends either in some $x_l$ or some  $y_k$ and each vertex
carries a factor  $(i\lambda)$, since the $4!$  cancels. Notice that many
different terms in the expansion of equation~(\ref{pert}) may give rise to the
same
Feynman diagram, through relabelling of lines or vertices. Let $N$ be the
number
of such equivalent terms. Assuming the diagram contains $m$  vertices as
above, and $q$ lines there is an explicit ${1\over m!}$
in the perturbative expansion and an explicit ${1\over q!}$ in the expansion of
$\exp(iZ_0[J])$. Then the combinatorial factor ${N\over m!q!}$  is called the
symmetry factor of the Feynman graph. In simple situations it is equal to 1
but may be smaller, when permutation of lines or vertices do not
correspond to different terms.

Finally the perturbative expansion of  $G(x_1,\cdots,x_n)$ at order $m$
is obtained by summing over all Feynman graphs with $m$ vertices,
each with its symmetry factor, and integrating over the variables $y_k$.
Notice that such graphs in general contain loops, i.e., are not of tree type.
In fact it is easy to see that there are exactly $m$ loops at order  $m$.

Unfortunately some diagrams with loops produce divergent results when
integrated over variables  $y_k$. This happens because $\Delta_F(x)$ is
singular for   $x=0$. This is cured by renormalization, as will be discussed
in R. Stora's lectures. One then gets a renormalized perturbation expansion
which is generally a divergent power series. In some particularly simple
situations this series can be resummed by appropriate methods. In practice
the first few terms of the perturbative expansion are considered as an accurate
representation of the ``exact'' quantum result. The classical limit
$\hbar\to 0$ reduces to the sum of tree graphs, i.e., graphs with no loop.

Finally we must mention that the above perturbative scheme breaks down
in the case of Gauge theories (which is relevant for the Standard model)
because the quadratic part of the Lagrangian happens to be non invertible,
precisely because of the symmetry of the theory. This means that
propagators do not exist, and the solution is to break the symmetry by
the addition of a  ``gauge--breaking term''. In turn we then have to add
new fields in the theory called ``Faddeev--Popov'' fields in order to
recover essentially the gauge symmetry in the form of a new symmetry,
mixing bosonic and fermionic fields called BRS symmetry. This symmetry
allows to show that the physical content of the theory is independent
of the gauge breaking term, so that a solid foundation for the
perturbative expansion is regained. This will be discussed in R. Stora's
lecture.

\section{The effective action}
\setcounter{equation}{0}

\subsection{Generating function of connected Green's functions}

We have already introduced (formally) the generating function
$W[J]=\int{\cal D}\phi\,\exp(iS_J)$ which
produces all Green's functions under functional differentiation with respect
to the external source  $J$. We have also seen that in the free case
$W[J]=\exp\,iZ[J]$. We shall see that such a formula holds true in
general and that $Z[J]$ is the generating function of connected Green's
function.
By a  connected Green's function we
mean the sum of connected Feynman diagrams, i.e., diagrams that cannot be
written as a product of disjoint parts (in particular, such diagrams do not
contain pure vacuum parts, i.e., each line is connected to at least one of the
terminals $x_k$).
For example, a disconnected four point function is of the form:
\par\vskip .5cm \hfil
\setlength{\unitlength}{0.005in}%
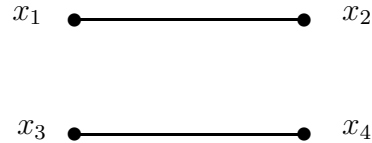
\begin{picture}(345,136)(95,593)
\vspace{4in}
\thicklines
\put(160,720){\circle*{14}}
\put(400,720){\circle*{14}}
\put(160,600){\circle*{14}}
\put(400,600){\circle*{14}}
\put(160,720){\line( 1, 0){240}}
\put(160,600){\line( 1, 0){240}}
\put( 95,720){\makebox(0,0)[lb]{\raisebox{0pt}[0pt][0pt]{$ x_1$}}}
\put(440,720){\makebox(0,0)[lb]{\raisebox{0pt}[0pt][0pt]{$ x_2$}}}
\put(100,600){\makebox(0,0)[lb]{\raisebox{0pt}[0pt][0pt]{$ x_3$}}}
\put(440,600){\makebox(0,0)[lb]{\raisebox{0pt}[0pt][0pt]{$ x_4$}}}
\end{picture}\par\vskip .5cm\noindent
Obviously in the free situation the only connected diagram is the two point
diagram:
\par\vskip .5cm \hfil
\setlength{\unitlength}{0.005in}%
\begin{picture}(360,16)(95,713)
\thicklines
\put(160,720){\circle*{14}}
\put(400,720){\circle*{14}}
\put(160,720){\line( 1, 0){240}}
\put( 95,720){\makebox(0,0)[lb]{\raisebox{0pt}[0pt][0pt]{$x_1$}}}
\put(440,720){\makebox(0,0)[lb]{\raisebox{0pt}[0pt][0pt]{$x_2$}}}
\end{picture}\par\vskip .5cm\noindent
and this corresponds to the fact that, in this case,
$Z[J]=-{1\over 2}\int J\Delta_F J$.

In the interacting case one can consider the following connected four point
functions:
\par \hfil
\setlength{\unitlength}{0.005in}%
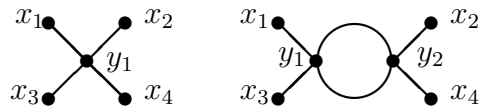
\begin{picture}(460,134)(80,675)
\thicklines
\put(120,760){\circle*{14}}
\put(160,720){\circle*{14}}
\put(200,680){\circle*{14}}
\put(120,680){\circle*{14}}
\put(200,760){\circle*{14}}
\put(360,760){\circle*{14}}
\put(360,680){\circle*{14}}
\put(400,720){\circle*{14}}
\put(480,720){\circle*{14}}
\put(520,760){\circle*{14}}
\put(520,680){\circle*{14}}
\put(440,720){\circle{80}}
\put(120,760){\line( 1,-1){ 80}}
\put(200,680){\line(-1, 1){ 80}}
\put(120,680){\line( 1, 1){ 80}}
\put(360,760){\line( 1,-1){ 40}}
\put(360,680){\line( 1, 1){ 40}}
\put(520,760){\line(-1,-1){ 40}}
\put(480,720){\line( 1,-1){ 40}}
\put( 85,760){\makebox(0,0)[lb]{\raisebox{0pt}[0pt][0pt]{$ x_1$}}}
\put(220,760){\makebox(0,0)[lb]{\raisebox{0pt}[0pt][0pt]{$x_2$}}}
\put( 80,680){\makebox(0,0)[lb]{\raisebox{0pt}[0pt][0pt]{$ x_3$}}}
\put(220,680){\makebox(0,0)[lb]{\raisebox{0pt}[0pt][0pt]{$ x_4$}}}
\put(180,715){\makebox(0,0)[lb]{\raisebox{0pt}[0pt][0pt]{$ y_1$}}}
\put(320,760){\makebox(0,0)[lb]{\raisebox{0pt}[0pt][0pt]{$ x_1$}}}
\put(540,760){\makebox(0,0)[lb]{\raisebox{0pt}[0pt][0pt]{$ x_2$}}}
\put(320,680){\makebox(0,0)[lb]{\raisebox{0pt}[0pt][0pt]{$ x_3$}}}
\put(540,680){\makebox(0,0)[lb]{\raisebox{0pt}[0pt][0pt]{$ x_4$}}}
\put(360,720){\makebox(0,0)[lb]{\raisebox{0pt}[0pt][0pt]{$ y_1$}}}
\put(505,720){\makebox(0,0)[lb]{\raisebox{0pt}[0pt][0pt]{$ y_2$}}}
\end{picture}\par\vskip .5cm\noindent
and so on.

Let us show that $Z[J]$  indeed generates connected Green's functions.
First, vacuum diagrams are removed by considering the quotient:
$${ \int {\cal D}\phi\, \phi_1\cdots\phi_n\,e^{iS_J}\over
\int {\cal D}\phi\,e^{iS_J}}$$
But setting  $W[J]=\exp\,iZ[J]$, a variation $\delta J_i$  gives rise to:
$$\delta W=i\delta Z \,W=\int {\cal D}\phi\, i\delta J_i\,\phi_i
\,e^{iS_J}$$
so that the above quotient is exactly for n=1 ($i$ collects all
indices, including space--time position):
\begin{equation}
{\delta Z\over\delta J_i}\equiv\Phi_i={1\over W}
\int {\cal D}\phi\, \phi_i\,e^{iS_J}\label{phic}
\end{equation}

Hence, in the limit $J\to 0$  is the vacuum expectation value of the
field  $\phi_i$, which may very well be equal to some constant
(by translation invariance) value $v_i$
(the situation with  $v_i\neq 0$ happens to be important in the
Standard model). We
then define for any $J$ the field $\bar{\phi}_i=\phi_i-v_i$
such that $<\phi_i>=0$  when $J=0$, i.e., $\bar{\phi_i}$  has no tadpole. Then
we
have:
\proclaim Theorem.
Functional differentiation of $Z[J]$
$$\left.{\delta^n Z[J]\over\delta J_1(x_1)\cdots\delta J_n(x_n)}
\right|_{J=0}=(i)^{n-1}<T\left(\bar{\phi}_1(x_1)\cdots\bar{\phi}_n(x_n)
\right)>^c$$
yields connected expectation values of time ordered
product of fields (denoted by the subscript $c$).
In particular the connected one point function vanishes.

\noindent{\it Proof.}  For any $J$  let us denote:
$$<\bar{\phi}_k>^c_J={1\over W[J]}\int {\cal D}\phi\,
\bar{\phi}_k\,e^{iS_J}=\Phi_k-v_k$$
$$<\bar{\phi}_k\bar{\phi}_l>_J={1\over W[J]}\int {\cal D}\phi\,
\bar{\phi}_k\bar{\phi}_l\,e^{iS_J}$$
Then we have:
\begin{eqnarray*}
{\delta\over\delta J_l}<\bar{\phi}_k>^c_J &=&
-i \Phi_l <\bar{\phi}_k>^c_J+<\bar{\phi}_k i(\bar{\phi}_l+v_l)>_J\\
&=& i<\bar{\phi}_k\bar{\phi}_l>_J-i<\bar{\phi}_k>^c_J<\bar{\phi}_l>^c_J
\end{eqnarray*}
We shall denote (skipping the $T$ symbol for brevity):
\begin{equation}
<\bar{\phi}_k\bar{\phi}_l>^c_J=
<\bar{\phi}_k\bar{\phi}_l>_J-<\bar{\phi}_k>^c_J<\bar{\phi}_l>^c_J
\label{recur}
\end{equation}
and we shall show that this two point function is indeed connected, so
that we get:
\begin{equation}
{\delta\over\delta J_l}<\bar{\phi}_k>^c_J=i<\bar{\phi}_k\bar{\phi}_l>^c_J
\label{princ}
\end{equation}
Notice that $<\bar{\phi}_k\bar{\phi}_l>^c_J$ is obtained by removing from
$<\bar{\phi}_k\bar{\phi}_l>_J$  the possible disconnected parts. One
then has to worry about the combinatorial factors, i.e., the symmetry factors.

Let us consider a disconnected graph that can be separated into two parts with
respectively $m_1$ loops, $q_1$ lines, weight $N_1$   (number of identical
terms in the perturbative expansion), and $m_2$, $q_2$, $N_2$. Then each
component occurs with combinatorial factor ${N_i\over m_i!q_i!}$ while
the disconnected graph will receive a factor ${N\over (m_1+m_2)!(q_1+q_2)!}$.
Each permutation of lines or vertices between the two components corresponds
to a different term in the perturbative expansion, precisely because the parts
are disjoint. The number of such moves is ${(m_1+m_2)!\over m_1!m_2!}
{(q_1+q_2)!\over q_1!q_2!}$. Multiplying by the number $N_1N_2$ of identical
terms in each connected part we get the total weight $N$, and we see that
the combinatorial factor of the disconnected graph is just the product of
the combinatorial factors of its parts. Obviously a similar reasoning works
for any number of parts. Finally the connected two point function is
obtained by removing from the ordinary two point function the disconnected
graphs, correctly counted, as it should be.

We can now proceed to the next inductive step.
\begin{eqnarray*}
{\delta\over\delta J_j}<\bar{\phi}_k\bar{\phi}_l>^c_J &=&
-i\Phi_j <\bar{\phi}_k\bar{\phi}_l>_J\\
&&+<\bar{\phi}_k\bar{\phi}_l i (\bar{\phi}_j+v_j)>_J\\
&&-i<\bar{\phi}_k\bar{\phi}_j>^c_J<\bar{\phi}_l>^c_J\\
&&-i<\bar{\phi}_l\bar{\phi}_j>^c_J<\bar{\phi}_k>^c_J
\end{eqnarray*}
Since we have by equation~(\ref{recur})
$$(-i\Phi_j +iv_j)<\bar{\phi}_k\bar{\phi}_l>_J=
-i<\bar{\phi}_j>^c_J\left(<\bar{\phi}_k\bar{\phi}_l>^c_J
+<\bar{\phi}_k>^c_J<\bar{\phi}_l>^c_J\right)$$
we get the equation similar to equation~(\ref{princ}):
$$ {\delta\over\delta J_j}<\bar{\phi}_k\bar{\phi}_l>^c_J=
i<\bar{\phi}_j\bar{\phi}_k\bar{\phi}_l>^c_J$$
with the definition similar to equation~(\ref{recur}):
\begin{eqnarray*}
<\bar{\phi}_j\bar{\phi}_k\bar{\phi}_l>^c_J &=&
<\bar{\phi}_j\bar{\phi}_k\bar{\phi}_l>_J\\
&&- <\bar{\phi}_j>^c_J <\bar{\phi}_k\bar{\phi}_l>^c_J
  -<\bar{\phi}_k>^c_J <\bar{\phi}_l\bar{\phi}_j>^c_J\\
&&-<\bar{\phi}_l>^c_J <\bar{\phi}_j\bar{\phi}_k>^c_J
 -<\bar{\phi}_j>^c_J<\bar{\phi}_k>^c_J<\bar{\phi}_l>^c_J
\end{eqnarray*}
By the preceding arguments it is obvious that this is the connected
three point function, and that this computation extends similarly to the
$n$--point function, thereby completing the proof of the theorem.

\subsection{One--particle--irreducible graphs}

We now proceed to define the generating function of one particle
irreducible graphs which turns out to be the {\em effective action},
that is the generalization of the
classical action at the quantum level. It is simply the Legendre transform
of the generating function $Z[J]$ of connected graphs, in the following way.
We have defined $\delta Z /\delta J_i=\Phi_i[J]$ as the mean value of
$\phi_i$ under path integration. Let us invert this relation to express
$J_i=J_i[\Phi]$ and form:
\begin{equation}
\Gamma[\Phi]=Z[J]-\sum_i J_i \Phi_i \label{gam}
\end{equation}
(more precisely the summation on $i$ means $\sum_i \int d^4x\, J_i(x)\Phi_i(x)$
).

Under variations $\delta J_i$ producing some $\delta \Phi_j$ we get:
$$\delta \Gamma=\sum_i {\delta Z\over\delta J_i}\delta J_i
-\sum_i \Phi_i\delta J_i-\sum_i J_i\delta\Phi_i=-\sum_i J_i\delta\Phi_i$$
so that:
$$J_i=-{\delta \Gamma\over\delta\Phi_i}$$
In particular, for $J\to 0$ all $\Phi_i$'s go to their vacuum expectation
values $v_i$ and we see that:
$$\left. {\delta \Gamma\over\delta\Phi}\right|_{\Phi=v}=-J=0$$
Hence $v$ is {\em the value of $\Phi$ which extremizes $\Gamma$} as in the
classical case with respect to the classical action.

Moreover consider the limit $\hbar \to 0$ in:
$$W[J]=e^{{i\over\hbar}Z[J]}=\int {\cal D}\phi \, e^{{i\over\hbar}S_J}$$
In this classical limit, everything reduces to the  action evaluated on the
classical solution so we have $Z[J]=S_J[\phi_{cl}]+O(\hbar)$ (where
$\phi_{cl}$ extremizes $S_J$) hence $Z[J]=S[\phi_{cl}]+J\phi_{cl}$. By
definition of $\phi_{cl}$ this quantity is stationary when $\phi_{cl}$ is
varied with $J$ fixed, so that only the explicit $J$ term contributes in:
$$\Phi={\delta Z\over \delta J}=\phi_{cl}$$
hence we get finally:
$$\Gamma[\Phi]=Z[J]-J\Phi=S[\Phi]$$

We have shown that $\Gamma[\Phi]=S[\Phi]$ at order 0 in $\hbar$, so that
$\Gamma[\Phi]$ is the quantum generalization of the classical action. In
general the symmetries present in $S$ will remain as such in $\Gamma$, this
being notably the case for the BRS symmetry, when the path
integration measure ${\cal D}\phi$ is formally invariant under the
considered symmetry. Nevertheless problems associated with the
divergences of quantum field theory may break this nice scheme.
In such cases the lack of symmetry of the effective action is
called an ``anomaly''. Of course even the first
(one loop) correction $\Gamma^1[\Phi]$ is completely non local, and
moreover divergent, but fortunately the divergent part is local and can be
absorbed in a redefinition of the classical $\Gamma^0$. This can be done at
all orders consistently in a renormalizable theory. An example of a direct
computation of $\Gamma^1$ can be found in Coleman--Weinberg~\cite{CW}, in a
simple situation, and more extensive computations along the same way
have been performed by S. Weinberg for the Standard model~\cite{W2}.

So it is nice to be able to compute $\Gamma[\Phi]$ by summing over
one--particle irreducible graphs. By this we mean graphs that cannot
be decomposed into disjoint parts by cutting one line. We begin by considering
propagators. In the free situation we have seen that $Z[J]=-1/2 \sum
J_i\Delta_{ij}J_j $ so that:
$${\delta^2 Z\over\delta J_i\delta J_j}=-\Delta_{ij}=
i<\bar{\phi}_i\bar{\phi}_j>^c$$
and $\Delta_{ij}$ is the ``bare'' propagator. In the interacting situation
we similarly define the ``dressed'' propagator $\Delta'_{ij}$ by:
\begin{equation}
{\delta^2 Z\over\delta J_i\delta J_j}=-\Delta'_{ij}\label{dres}
\end{equation}
As a matter of fact $\Delta'_{ij}$ is the sum of connected graphs leading
from $i$ to $j$ (the bubbles below mean two point diagrams which
cannot be separated into two parts by cutting a line):
\par\vskip .5cm \hfil
\setlength{\unitlength}{0.005in}%
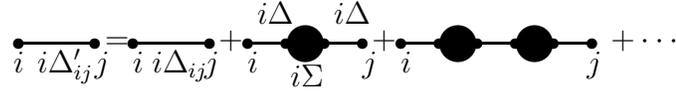
\begin{picture}(625,74)(35,715)
\thicklines
\put( 40,760){\circle*{10}}
\put(120,760){\circle*{10}}
\put(160,760){\circle*{10}}
\put(240,760){\circle*{10}}
\put(280,760){\circle*{10}}
\put(320,760){\circle*{10}}
\put(360,760){\circle*{10}}
\put(400,760){\circle*{10}}
\put(440,760){\circle*{10}}
\put(480,760){\circle*{10}}
\put(520,760){\circle*{10}}
\put(560,760){\circle*{10}}
\put(600,760){\circle*{10}}
\put(640,760){\circle*{10}}
\put(340,760){\circle*{40}}
\put(500,760){\circle*{40}}
\put(580,760){\circle*{40}}
\put( 40,760){\line( 1, 0){ 80}}
\put(120,760){\line(-1, 0){ 80}}
\put(160,760){\line( 1, 0){ 80}}
\put(280,760){\line( 1, 0){ 40}}
\put(360,760){\line( 1, 0){ 40}}
\put(440,760){\line( 1, 0){ 40}}
\put(520,760){\line( 1, 0){ 40}}
\put(600,760){\line( 1, 0){ 40}}
\put( 35,730){\makebox(0,0)[lb]{\raisebox{0pt}[0pt][0pt]{$ i$}}}
\put( 60,730){\makebox(0,0)[lb]{\raisebox{0pt}[0pt][0pt]{$i\Delta'_{ij}$}}}
\put(120,730){\makebox(0,0)[lb]{\raisebox{0pt}[0pt][0pt]{$j$}}}
\put(160,730){\makebox(0,0)[lb]{\raisebox{0pt}[0pt][0pt]{$i$}}}
\put(180,730){\makebox(0,0)[lb]{\raisebox{0pt}[0pt][0pt]{$i\Delta_{ij}$}}}
\put(235,730){\makebox(0,0)[lb]{\raisebox{0pt}[0pt][0pt]{$j$}}}
\put(280,730){\makebox(0,0)[lb]{\raisebox{0pt}[0pt][0pt]{$i$}}}
\put(325,715){\makebox(0,0)[lb]{\raisebox{0pt}[0pt][0pt]{$i\Sigma$}}}
\put(290,780){\makebox(0,0)[lb]{\raisebox{0pt}[0pt][0pt]{$i\Delta$}}}
\put(370,780){\makebox(0,0)[lb]{\raisebox{0pt}[0pt][0pt]{$i\Delta$}}}
\put(400,730){\makebox(0,0)[lb]{\raisebox{0pt}[0pt][0pt]{$j$}}}
\put(440,730){\makebox(0,0)[lb]{\raisebox{0pt}[0pt][0pt]{$i$}}}
\put(635,730){\makebox(0,0)[lb]{\raisebox{0pt}[0pt][0pt]{$j$}}}
\put(120,755){\makebox(0,0)[lb]{\raisebox{0pt}[0pt][0pt]{ $=$}}}
\put(250,755){\makebox(0,0)[lb]{\raisebox{0pt}[0pt][0pt]{$+$}}}
\put(410,755){\makebox(0,0)[lb]{\raisebox{0pt}[0pt][0pt]{$+$}}}
\put(660,755){\makebox(0,0)[lb]{\raisebox{0pt}[0pt][0pt]{$+\cdots$}}}
\end{picture}\par\vskip .5cm\noindent
This is a geometric series which sums up to $1 / (\Delta^{-1}+\Sigma)$.
When $\Delta =1/(k^2-\mu^2)$ we get $\Delta'=1/(k^2-\mu^2+\Sigma(p^2))$
and $\Sigma(p^2)$ describes the (one-particle-irreducible)
quantum corrections to the propagator, which
are called the self--energy corrections. Essentially this shifts the
position of the pole of the propagator, i.e., the mass of the particle,
and introduces a broadening of this pole, through the imaginary part
of $\Sigma(\mu^2)$.

Similarly to equation~(\ref{dres}) we define:
$${\delta^2 \Gamma\over\delta\Phi_i\delta\Phi_j}=X_{ij}$$
and we have since $\delta Z/\delta J_i=\Phi_i$:
$$\sum_j {\delta^2Z\over\delta J_i\delta J_j}{\delta J_j\over\delta\Phi_k}
={\delta\Phi_i\over\delta\Phi_k}=\delta_{ik}$$
Similarly $J_j=-\delta\Gamma/\delta\Phi_j$ hence
$\delta J_j/\delta\Phi_k=-\delta^2\Gamma/\delta\Phi_j\delta\Phi_k$ so that:
\begin{equation}
\sum_j \Delta'_{ij} X_{jk}=\delta_{ik}\label{inpro}
\end{equation}
Recalling that $\Delta'_{ij}$ is the dressed propagator while
$$\Gamma[\Phi]=\sum_{jk}{1\over 2}\Phi_jX_{jk}\Phi_k+O(\Phi^3)$$
we see that the dressed propagator is the inverse of the coefficient of the
quadratic term in the effective action. In particular in the classical limit
{\em the bare propagator is the inverse of the quadratic part of the
Lagrangian}, which may be seen as the lowest order 1PI contribution to
$\Gamma$. When interactions are taken into account, $X_{jk}$ is obtained
by adding to this $\Sigma_{jk}$ i.e., all 1PI two point functions.

Then one derives the relation~(\ref{inpro}) with respect to $J_k$:
$${\delta^3 Z\over \delta J_i\delta J_j\delta J_k}X_{jl}+
(-\Delta'_{ij}){\delta^3\Gamma\over\delta\Phi_j\delta\Phi_l\delta\Phi_m}
{\delta\Phi_m\over\delta J_k}=0$$
Noticing that $\delta\Phi_m/\delta J_k=-\Delta'_{km}$ and inverting
$X_{jl}$ with a $\Delta'$ we get:
$${\delta^3 Z\over \delta J_i\delta J_j\delta J_k}=
-\Delta'_{il}\Delta'_{jm}\Delta'_{kn}{\delta^3\Gamma\over
\delta\Phi_l\delta\Phi_m\delta\Phi_n}$$
Defining the 3--vertex $\Gamma^{(3)}_{lmn}$ as $\delta^3\Gamma/
\delta\Phi_l\delta\Phi_m\delta\Phi_n$ we see that:
\begin{equation}
<T(\bar{\phi}_i\bar{\phi}_j\bar{\phi}_k)>^c=
(i\Delta'_{il})(i\Delta'_{jm})(i\Delta'_{kn})(i\Gamma^{(3)}_{lmn})
\label{3vert}\end{equation}
This means that the connected three point function is obtained by attaching
dressed propagators to the dressed vertex $\Gamma^{(3)}_{lmn}$ which is
therefore the so--called amputated three point function. Obviously
there is no way to separate such a function into parts by cutting a line,
so the amputated vertex is 1PI.
\par\vskip .5cm \hfil
\setlength{\unitlength}{0.005in}%
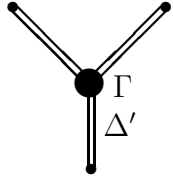
\begin{picture}(170,180)(115,585)
\thicklines
\put(120,760){\circle*{10}}
\put(280,760){\circle*{10}}
\put(200,680){\circle*{30}}
\put(202,590){\circle*{10}}
\put(120,755){\line( 1,-1){ 80}}
\put(200,675){\line(-1, 1){ 80}}
\put(125,760){\line( 1,-1){ 75}}
\put(275,760){\line(-1,-1){ 75}}
\put(280,755){\line(-1,-1){ 75}}
\put(199,677){\line( 0,-1){ 84}}
\put(205,669){\line( 0,-1){ 76}}
\put(225,665){\makebox(0,0)[lb]{\raisebox{0pt}[0pt][0pt]{$\Gamma$}}}
\put(215,625){\makebox(0,0)[lb]{\raisebox{0pt}[0pt][0pt]{$\Delta'$}}}
\end{picture}\par\vskip .5cm

Taking as an example a theory with a coupling ${\lambda\over 3!}\phi^3$
the lowest order value of $\Gamma^{(3)}$ is precisely $\lambda$ and at the next
order we have the 1PI diagram:
\par\vskip .5cm \hfil
\setlength{\unitlength}{0.005in}%
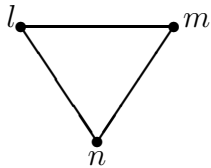
\begin{picture}(240,200)(120,600)
\thicklines
\put(160,760){\circle*{10}}
\put(320,760){\circle*{10}}
\put(240,640){\circle*{10}}
\put(160,760){\line( 2,-3){ 80}}
\put(160,760){\line( 1, 0){160}}
\put(320,760){\line(-2,-3){ 80}}
\put(145,760){\makebox(0,0)[lb]{\raisebox{0pt}[0pt][0pt]{$l$}}}
\put(330,760){\makebox(0,0)[lb]{\raisebox{0pt}[0pt][0pt]{$m$}}}
\put(230,615){\makebox(0,0)[lb]{\raisebox{0pt}[0pt][0pt]{$n$}}}
\end{picture}\par\vskip .5cm\noindent

We can now state the:
\proclaim Theorem.
Functional differentiation of the effective action $\Gamma[\Phi]$ yields
the one--particle--irreducible graphs.

\noindent{\it Proof.} Inductively assume that:
$${1\over i^{n-1}}{\delta^n Z[J]\over\delta J_i \delta J_j\cdots}=
(i\Delta'_{il})(i\Delta'_{jm})\cdots i{\delta^n \Gamma[\Phi]\over
\delta\Phi_l \delta\Phi_m\cdots}+\mbox{~1--part.~red.~graphs}$$
and differentiate with:
$${1\over i}{\delta\over\delta J_k}=(i\Delta'_{kr}){\delta\over\delta\Phi_r}$$
Either this derivation acts on some $\delta^m\Gamma/\delta\Phi_\alpha\cdots$
and produces \break\hfil
$(i\Delta'_{kr})\delta^{m+1}\Gamma/
\delta\Phi_r\delta\Phi_\alpha\cdots$ i.e., the external leg $k$ is attached
to an $m$--vertex producing a $(m+1)$--vertex (the corresponding graph
will be one--particle--reducible except if $m=n$) or the derivation acts
on some $(i\Delta'_{m\alpha})$. Then as we have seen:
$${1\over i}{\delta\over\delta J_k}(i\Delta'_{m\alpha})=
-{\delta^3 Z\over\delta J_k\delta J_m\delta J_\alpha}=
(i\Delta'_{ka})(i\Delta'_{mb})(i\Delta'_{\alpha c})\Gamma^{(3)}_{abc}$$
This means that the leg $k$ gets attached in the following way:
\par\vskip .5cm \hfil
\setlength{\unitlength}{0.005in}%
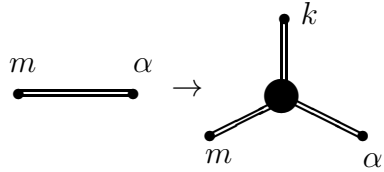
\begin{picture}(375,160)(70,645)
\thicklines
\put( 80,722){\circle*{10}}
\put(200,722){\circle*{10}}
\put(280,678){\circle*{10}}
\put(440,678){\circle*{10}}
\put(358,800){\circle*{10}}
\put(355,720){\circle*{36}}
\put( 80,720){\line( 1, 0){120}}
\put( 80,725){\line( 1, 0){120}}
\put(280,680){\line( 2, 1){ 80}}
\put(360,720){\line( 2,-1){ 80}}
\put(360,800){\line( 0,-1){ 80}}
\put(280,675){\line( 2, 1){ 80}}
\put(360,715){\line( 2,-1){ 80}}
\put(355,800){\line( 0,-1){ 80}}
\put(240,720){\makebox(0,0)[lb]{\raisebox{0pt}[0pt][0pt]{$\to$}}}
\put( 70,745){\makebox(0,0)[lb]{\raisebox{0pt}[0pt][0pt]{$ m$}}}
\put(200,745){\makebox(0,0)[lb]{\raisebox{0pt}[0pt][0pt]{$\alpha$}}}
\put(275,650){\makebox(0,0)[lb]{\raisebox{0pt}[0pt][0pt]{$m$}}}
\put(440,645){\makebox(0,0)[lb]{\raisebox{0pt}[0pt][0pt]{$\alpha$}}}
\put(375,795){\makebox(0,0)[lb]{\raisebox{0pt}[0pt][0pt]{$k$}}}
\end{picture}
\par\vskip .5cm\noindent
This always give a one--particle--reducible graph. For example, starting
from the above three point function we get the four point function as:
\par\vskip .5cm \hfil
\setlength{\unitlength}{0.005in}%
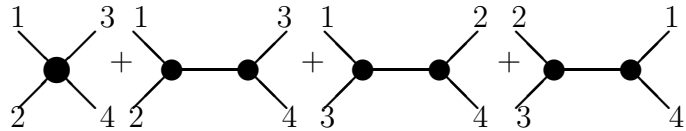
\begin{picture}(700,109)(20,705)
\thicklines
\put( 80,760){\circle*{28}}
\put(200,760){\circle*{22}}
\put(280,760){\circle*{22}}
\put(400,760){\circle*{22}}
\put(480,760){\circle*{22}}
\put(600,760){\circle*{22}}
\put(680,760){\circle*{22}}
\put( 40,800){\line( 1,-1){ 80}}
\put(120,800){\line(-1,-1){ 80}}
\put(160,800){\line( 1,-1){ 40}}
\put(200,760){\line(-1,-1){ 40}}
\put(200,760){\line( 1, 0){ 80}}
\put(280,760){\line( 1, 1){ 40}}
\put(280,760){\line( 1,-1){ 40}}
\put(360,800){\line( 1,-1){ 40}}
\put(400,760){\line(-1,-1){ 40}}
\put(480,760){\line( 1, 1){ 40}}
\put(480,760){\line( 1,-1){ 40}}
\put(400,760){\line( 1, 0){ 80}}
\put(560,800){\line( 1,-1){ 40}}
\put(560,720){\line( 1, 1){ 40}}
\put(600,760){\line( 1, 0){ 80}}
\put(680,760){\line( 1, 1){ 40}}
\put(680,760){\line( 1,-1){ 40}}
\put( 20,805){\makebox(0,0)[lb]{\raisebox{0pt}[0pt][0pt]{ 1}}}
\put(125,805){\makebox(0,0)[lb]{\raisebox{0pt}[0pt][0pt]{3}}}
\put( 20,700){\makebox(0,0)[lb]{\raisebox{0pt}[0pt][0pt]{ 2}}}
\put(125,700){\makebox(0,0)[lb]{\raisebox{0pt}[0pt][0pt]{4}}}
\put(135,760){\makebox(0,0)[lb]{\raisebox{0pt}[0pt][0pt]{+}}}
\put(160,805){\makebox(0,0)[lb]{\raisebox{0pt}[0pt][0pt]{1}}}
\put(310,805){\makebox(0,0)[lb]{\raisebox{0pt}[0pt][0pt]{3}}}
\put(155,700){\makebox(0,0)[lb]{\raisebox{0pt}[0pt][0pt]{2}}}
\put(315,700){\makebox(0,0)[lb]{\raisebox{0pt}[0pt][0pt]{4}}}
\put(355,805){\makebox(0,0)[lb]{\raisebox{0pt}[0pt][0pt]{1}}}
\put(515,805){\makebox(0,0)[lb]{\raisebox{0pt}[0pt][0pt]{2}}}
\put(355,700){\makebox(0,0)[lb]{\raisebox{0pt}[0pt][0pt]{3}}}
\put(515,700){\makebox(0,0)[lb]{\raisebox{0pt}[0pt][0pt]{4}}}
\put(335,760){\makebox(0,0)[lb]{\raisebox{0pt}[0pt][0pt]{+}}}
\put(555,805){\makebox(0,0)[lb]{\raisebox{0pt}[0pt][0pt]{2}}}
\put(715,805){\makebox(0,0)[lb]{\raisebox{0pt}[0pt][0pt]{1}}}
\put(560,700){\makebox(0,0)[lb]{\raisebox{0pt}[0pt][0pt]{3}}}
\put(720,700){\makebox(0,0)[lb]{\raisebox{0pt}[0pt][0pt]{4}}}
\put(540,760){\makebox(0,0)[lb]{\raisebox{0pt}[0pt][0pt]{+}}}
\end{picture}
\par\vskip .5cm\noindent
The five point function is obtained similarly by attaching a new leg
to the four external legs, moreover we get new diagrams by attaching
the new leg to the internal dressed propagator as in:
\par\vskip .5cm \hfil
\setlength{\unitlength}{0.005in}%
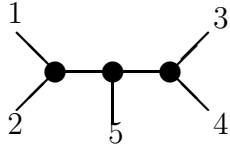
\begin{picture}(215,134)(30,685)
\thicklines
\put( 80,760){\circle*{22}}
\put(200,760){\circle*{22}}
\put(140,760){\circle*{22}}
\put( 40,800){\line( 1,-1){ 40}}
\put( 40,720){\line( 1, 1){ 40}}
\put( 80,760){\line( 1, 0){120}}
\put(200,760){\line( 1, 1){ 40}}
\put(200,760){\line( 1,-1){ 40}}
\put(140,760){\line( 0,-1){ 55}}
\put( 30,810){\makebox(0,0)[lb]{\raisebox{0pt}[0pt][0pt]{1}}}
\put( 30,695){\makebox(0,0)[lb]{\raisebox{0pt}[0pt][0pt]{2}}}
\put(245,805){\makebox(0,0)[lb]{\raisebox{0pt}[0pt][0pt]{3}}}
\put(245,695){\makebox(0,0)[lb]{\raisebox{0pt}[0pt][0pt]{4}}}
\put(135,685){\makebox(0,0)[lb]{\raisebox{0pt}[0pt][0pt]{5}}}
\end{picture}
\par\vskip .5cm\noindent
It is now clear that the recursion is verified at order $(n+1)$ and
moreover that the remaining one--particle--reducible graphs form
{\em all the tree graphs} that can be constructed with the irreducible
vertices $\Gamma^{(m)}$.

This means that the full quantum theory of the fields $\phi_i$ with
the Lagrangian $L[\phi]$ is the same as the classical limit of the
theory of the fields $\Phi_i$ with the effective action $\Gamma[\Phi]$,
a fact which is easily verified by noting that for $a\to 0$ the path integral
$\int {\cal D}\Phi \exp {i\over a}\left\{ \Gamma[\Phi]+J\Phi\right\}$
becomes equal to $\exp {i\over a}\left\{ \Gamma[\Phi_S]+J\Phi_S\right\}$
where $\Phi_S$ is the stationary value, i.e., such that
$J=-\delta\Gamma/\delta\Phi$ hence equal to $\exp {i\over a}Z[J]$.

Of course to fully justify these considerations one still has to
show that the combinatorial weights of the diagrams match correctly.
The argument is similar as in the case of connected diagrams.
Assume that a one--particle--reducible graph is obtained by joining
two parts with respectively $m_i$ lines $q_i$ vertices and weight $N_i$
with a dressed propagator. The resulting graph has $(m_1+m_2+1)$ lines,
$(q_1+q_2)$ vertices and weight $N$. Moves that contribute
to $N$ are permutation of vertices between the two parts whose
number is ${(q_1+q_2)!\over q_1!q_2!}$. All permutation of lines
between the two parts, and exchange of the line which connects the two parts
with one of the other correspond to different terms of the perturbative
expansion
due to the topology of the graph. The number of these moves is given by
dividing the number $(m_1+m_2+1)!$ of all permutation of lines by the
number $m_1!m_2!$ of permutations inside each part. Hence we get:
$$N=N_1 N_2 {(q_1+q_2)!\over q_1!q_2!}{ (m_1+m_2+1)!\over m_1!m_2!}$$
so that the symmetry number of the whole is the product of the
symmetry number of its parts. This completes the proof.

\section{Goldstone and Higgs mechanisms}
\setcounter{equation}{0}

\subsection{The Goldstone theorem}

The Goldstone theorem concerns a theory in which a global symmetry is
spontaneously broken. Let us take the simplest example of two scalar
fields $\sigma$ and $\pi$ with a $U(1)$ symmetry. The corresponding
Lagrangian density is:
$$L={1\over 2}\left( (\partial_\mu\sigma)^2+(\partial_\mu\pi)^2\right)
-{\mu^2\over 2}(\sigma^2+\pi^2)-{\lambda\over 4}(\sigma^2+\pi^2)^2$$
and the global symmetry corresponds to the rotation of fields:
$$\pmatrix{\sigma\cr\pi\cr}\to\pmatrix{\cos\theta&-\sin\theta\cr
\sin\theta&\cos\theta\cr}.\pmatrix{\sigma\cr\pi\cr}$$

Let us now assume that the mass $\mu$ has the ``unphysical'' value
such that $\mu^2 < 0$. Then the minimal value of the potential is given
by:
$$ \sigma^2+\pi^2=-{\mu^2\over\lambda}$$
In other words there is a whole circle of minima (due to the symmetry) and
one has to {\em choose} one of them as the starting point of the perturbative
expansion. This is called spontaneous symmetry breakdown. On the contrary for
physical values of the mass there is only one minimum at $\sigma=\pi=0$.
So let us take the minimum at $\sigma=\sigma_0$ and $\pi=0$ hence
$\sigma_0=\sqrt{-\mu^2/\lambda}$. One analyzes the theory by simply shifting
the fields according to $\sigma\to\sigma_0+\sigma$, $\pi\to\pi$, so that
the vacuum now corresponds to $\sigma=\pi=0$.

The Lagrangian density is immediately written using these variables:
$$L={1\over 2}\left( (\partial_\mu\sigma)^2+(\partial_\mu\pi)^2\right)
+\mu^2\sigma^2-\lambda\sigma_0\sigma(\sigma^2+\pi^2)
-{\lambda\over 4}(\sigma^2+\pi^2)^2$$
This means that at the classical level, the field $\sigma$ has a
(now physical) mass term $-2\mu^2$ while the field $\pi$ has mass 0.
This is the content of the Goldstone theorem. In the situation in which
a global symmetry is spontaneously broken there appear particles with
no mass called Goldstone particles. Of course in realistic models of
particle physics this may be useful in some circumstances (for example
the pion may be seen as a Goldstone particle in the symmetric limit) but
is more frequently considered catastrophic since light particles cannot
be concealed and relegated to some high energy region. So it is very important
that the Goldstone theorem remains true at the quantum level. This has been
proven
by Goldstone~\cite{Go} by using direct quantum mechanical arguments, and also
by
Jona--Lasinio~\cite{JL} using the formalism of the effective action.

The proof runs as follows, using the notations of the previous subsection. Let
us consider an infinitesimal transformation $\delta \phi_i(x)=t_{ij}\phi_j(x)$
under
which $S[\phi]$ is invariant. If one compensates with $\delta
J_i(x)=-t_{ji}J_j(x)$
the product $J_i\phi_i$ is invariant, hence $Z[J]$ and $\Gamma[\Phi]$ are also
invariant. This yields the equality:
$$-t_{ji}J_j=\delta J_i=\sum_k\int d^4x\, {\delta
J_i\over\delta\Phi_k(x)}\delta\Phi_k(x)$$
$$=-\sum_k\int
d^4x\,{\delta^2\Gamma\over\delta\Phi_i\delta\Phi_k(x)}t_{kl}\Phi_l(x)$$
So in the limit $J\to 0$ we get (notice that $\int d^4x$ yields the Fourier
coefficient
at zero momentum):
$$\sum_k X_{ik}({\rm momentum}=0). t_{kl}v_l=0$$
Hence all transformations that effectively move the vacuum expectation value
$v$
(or that break the vacuum) also produce eigenvectors of the inverse propagator
for
the eigenvalue 0 at zero momentum. This means that there are as many modes
developing a pole of the propagator at zero mass, i.e., Goldstone particles. In
general there will be an isotropy subgroup of the vacuum $v$ and the number of
the Goldstone particles will be equal to the dimension of the quotient of
the symmetry group by this isotropy subgroup.

Fortunately there is a way out of this situation when the symmetry is realized
as
a gauge symmetry, so that the Goldstone modes can be gauged away. This is the
Higgs mechanism.

\subsection{The Higgs mechanism}

We shall explain this mechanism using the above simple model and denoting
$\phi=\sigma+i\pi$ so that the $U(1)$ symmetry reads $\phi\to e^{i\theta}\phi$.
This symmetry can be localized, i.e., one can take for $\theta$ a function of
the space--time point $x$ by adding a gauge field $A_\mu(x)$ and introducing
the
{\em covariant derivative} $D_\mu\phi=\partial_\mu\phi-ieA_\mu\phi$. Then
under the above symmetry, $D_\mu\phi\to e^{i\theta} D_\mu\phi$ if $A_\mu$
transforms according to $A_\mu\to A_\mu+{1\over e}\partial_\mu\theta$ (here
$e$ is the ``charge'' or the coupling constant between $A_\mu$ and $\phi$,
hence is the same for different multiplets). Using the ``field--strength''
$F_{\mu\nu}$ which is invariant under gauge transformation the Lagrangian
density reads:
$$L=-{1\over 4}F_{\mu\nu}F^{\mu\nu}+(D_\mu\phi)^*(D^\mu\phi)-\mu^2\phi^*\phi
-\lambda(\phi^*\phi)^2$$
All this situation can  be generalized readily to any semisimple symmetry group
(then $F_{\mu\nu}$ is covariant) following Yang and Mills~\cite{YM}.

Still assuming unphysical mass one chooses $<\phi>_0=v/\sqrt{2}$ with a real
$v$
such that $v=\sqrt{-\mu^2/\lambda}$. Shifting as above and expanding there
seems to appear a Goldstone particle. In fact it can be gauged away as follows.
One can always parametrize $\phi$ as:
$$\phi=e^{i\xi/v}\;{v+\eta\over\sqrt{2}}$$
and perform the gauge transformation:
$$\phi\to e^{-i\xi/v}\phi={v+\eta\over\sqrt{2}}\quad
A_\mu\to A_\mu-{1\over ev}\partial_\mu\xi$$
Since $L$ is gauge invariant it now reads:
$$L=-{1\over 4}F_{\mu\nu}F^{\mu\nu}+{1\over 2}D_\mu (v+\eta)D^\mu(v+\eta)-
{\mu^2\over 2}(v+\eta)^2 -{\lambda\over 4}(v+\eta)^4$$
where of course the new $A_\mu$ has been used. Notice that $\xi$ has completely
disappeared from the theory and one recovers a coupling:
$${1\over 2}\left(D_\mu (v+\eta)\right)^2={1\over 2}\left(\partial_\mu
\eta\right)^2
+{1\over 2}e^2A_\mu A^\mu(v^2+2v\eta+\eta^2)$$
Replacing $v$ by its value we see that $\eta$ has acquired a mass term
$-2\mu^2$
as in the Goldstone situation but moreover the vector field $A_\mu$ which
seemed to be massless has also acquired a mass $ev$. Notice that this mass
can be adjusted by varying $v$ or equivalently $\lambda$, but then other
particles in the theory similarly coupled to $\phi$ will acquire a mass
proportional to their coupling constant. Hence the Higgs mechanism
succeeds in both giving a mass to the vector field and gauging away the
Goldstone particle. This is the main ingredient in the construction of
the Standard model.

\subsection{The Standard model}

We finally describe briefly the ``unified'' model of weak and electromagnetic
interactions due to Glashow Salam Weinberg~\cite{We}. The weak interactions
are notably responsible of the decay of the neutron or the muon, and were as
such
described traditionally by a four fermions interaction. Such a theory presents
numerous problems; so T.D. Lee and C.N. Yang long ago postulated the
existence of a heavy charged vector meson $W$ mediating the weak interaction.
\par\vskip .5cm \hfil
\setlength{\unitlength}{0.005in}%
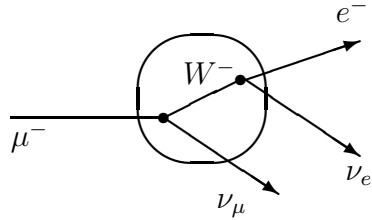
\begin{picture}(365,201)(80,545)
\thicklines
\put(280,660){\oval(134,134)}
\put(240,640){\circle*{10}}
\put(320,680){\circle*{10}}
\put( 80,640){\line( 1, 0){160}}
\put(325,680){\vector( 3,-2){120}}
\put(240,640){\vector( 3,-2){120}}
\put(325,680){\vector( 3, 1){120}}
\put(240,640){\line( 2, 1){ 80}}
\put( 80,610){\makebox(0,0)[lb]{\raisebox{0pt}[0pt][0pt]{$\mu^-$}}}
\put(420,740){\makebox(0,0)[lb]{\raisebox{0pt}[0pt][0pt]{$e^-$}}}
\put(295,545){\makebox(0,0)[lb]{\raisebox{0pt}[0pt][0pt]{$\nu_\mu$}}}
\put(430,575){\makebox(0,0)[lb]{\raisebox{0pt}[0pt][0pt]{$\nu_e$}}}
\put(260,675){\makebox(0,0)[lb]{\raisebox{0pt}[0pt][0pt]{$W^-$}}}
\end{picture}
\par\vskip .5cm\noindent
Unfortunately the theory of a massive vector field is also non
renormalizable so it was necessary to wait till the introduction of
the Higgs mechanism so as to get a consistent theory along these lines.
The simplest gauge theory involving gauge mesons $W^+$ and $W^-$ is based
on the group $SU(2)$ hence will also involve a neutral meson. As a matter
of fact, the above authors found that it was necessary to go to the next
simple situation, with a gauge group $SU(2)\times U(1)$ hence involving
two neutral mesons, one of them being the photon. This produced a
consistent framework for studying weak and electromagnetic interactions.

Since the gauge group is a direct product, there are two coupling constants
in the model, associated to a triplet $\vec{A}_\mu$ and a singlet $B_\mu$.
The corresponding covariant derivative is ($\vec{\sigma}$ are the Pauli
matrices):
$$D_\mu=\partial_\mu - ig \vec{A}_\mu.\vec{\sigma}-ig'B_\mu$$
while the Yang--Mills Lagrangian of the vector fields reads:
$$L_{YM}=-{1\over 4}B_{\mu\nu}B^{\mu\nu}-{1\over 4}
\vec{A}_{\mu\nu}\vec{A}^{\mu\nu}$$
where the field--strengths are defined as usual as:
$$B_{\mu\nu}=\partial_\mu B_\nu-\partial_\nu B_\mu \quad
\vec{A}_{\mu\nu}=\partial_\mu \vec{A}_\nu-\partial_\nu \vec{A}_\mu
+g \vec{A}_\mu\times \vec{A}_\nu$$

In order to give a mass to the appropriate vectors one introduces a
doublet of complex scalars, the Higgs field:
$$\phi=\pmatrix{\phi^+\cr\phi^0\cr}$$
where $\phi^+$ has electric charge $+1$ and $\phi^0$ has no electric charge.
One assumes that the potential is such that spontaneous symmetry
breakdown occurs so that the neutral component $\phi^0$ gets a
vacuum expectation value $v/\sqrt{2}$. Hence the Higgs field has to be
shifted around:
$$<\phi>={1\over\sqrt{2}}\pmatrix{0\cr v\cr}$$
when considering its Lagrangian:
$$L_{\rm Higgs}=(D_\mu \phi)^*(D^\mu \phi)+L_I(\phi^*\phi)$$
where $L_I$ is some  quadratic polynomial providing a mass
term and a coupling term for the Higgs. The action of  $SU(2)\times U(1)$ on
$\phi$ has an isotropy subgroup of dimension 1 hence one can gauge away
three out of the four real fields involved in $\phi$,  ending with just
one real electrically neutral field $h$ which is called {\em the}
Higgs field (what is really physical in the Higgs field). Hence one can
set
$$\phi={1\over\sqrt{2}}\pmatrix{0\cr v+h\cr}$$
in the Higgs Lagrangian, which then reads:
$$L_{\rm Higgs}={1\over 2}(\partial_\mu h)^2+
{1\over 4}g^2( v+h)^2W_\mu^+W^{\mu -}+{1\over 8}( v+h)^2
(g'B_\mu-gA_\mu^3)^2$$
where we have introduced as usual $W_\mu^\pm=(A_\mu^1\mp iA_\mu^2)/\sqrt{2}$.

One finds the mass matrix of the vectors by keeping the quadratic terms
$( v+h)^2\to v^2$ in the form
$${1\over 8}v^2\left\{(g'B_\mu-gA_\mu^3)^2+2g^2W_\mu^+W^{\mu -}\right\}$$
This means that the charged vector meson acquires a mass (or has a mass
term $M_W^2 \, W_\mu^+W^{\mu -}$)
\begin{equation}
M_W={1\over 2}gv=M_{W^+}=M_{W^-}
\end{equation}
while the neutral ones are still mixed. One diagonalizes the mass matrix
by performing a rotation of fields through the so--called Weinberg
angle $\theta_W$ defining:
\begin{eqnarray}
A_\mu &=& \cos\theta_W B_\mu +\sin\theta_W A_\mu^3 \nonumber\\
Z_\mu &=& \sin\theta_W B_\mu -\cos\theta_W A_\mu^3
\end{eqnarray}
Since this is a rotation it does not affect the form of the kinetic term
(invariance of the quadratic form) and one chooses $\theta_W$ so that
$(g'B_\mu-gA_\mu^3)$ be proportional to $Z_\mu$ hence:
\begin{equation}
\cos\theta_W={g\over\sqrt{g^2+g'^2}}\quad \sin\theta_W=
{g'\over\sqrt{g^2+g'^2}}
\end{equation}
Then obviously $Z_\mu$ acquires a mass (or has a mass term $M_Z^2\, Z_\mu
Z^\mu /2$)
\begin{equation}
M_Z={1\over 2}\sqrt{g^2+g'^2}\, v
\end{equation}
while $A_\mu$ remains massless and is identified to the photon. Of course
by developing the Yang--Mills Lagrangian $L_{YM}$ with these notations
one immediately finds trilinear and quadrilinear couplings between all
these vectors, including the correct coupling (with charge $e$: the
electronic charge) of the photon field $A_\mu$ with the charged bosons
$W_\mu^\pm$ while the neutral $Z_\mu$ remains uncoupled to $A_\mu$.
Notice that the previous equations lead to a relation between $M_W$
and $M_Z$:
$$\rho={M_W^2\over M_Z^2\cos^2\theta_W}=1$$
As it happens $\rho$ can be measured as the ratio of low energy interactions
of charged and neutral currents and the deviations of $\rho$ to one, due
to radiative corrections, are a sensitive test of the theory.

It now remains to introduce the fermions in the model. We shall present
the standard example of the electron and its neutrino. One first decomposes
the fermionic fields into chirality components:
$$L={1-\gamma^5\over 2}\quad R={1+\gamma^5\over 2}\quad
e_L=Le\quad e_R=Re\quad \nu_L=L\nu=\nu \quad \nu_R=0$$
and one affects the $L$ components to a doublet of $SU(2)$ with ``hypercharge''
$-1/2$ under the $U(1)$ while $e_R$ is a singlet of $SU(2)$ with
hypercharge $-1$. So the left doublet, on which $SU(2)$ acts is:
$$\psi_L=\pmatrix{\nu_L\cr e_L\cr}$$
and considering the covariant derivatives one immediately sees that the
electron is correctly coupled to the photon while the neutrino is neutral:
$$L_F=\bar{\psi}_L i\gamma^\mu D_\mu \psi_L+ \bar{e}_R i\gamma^\mu D_\mu e_R$$
reads after substituting the above definitions:
\begin{equation}
L_F=\bar{e}(i\not\!\partial-e\not\!\!
A)e+\bar{\nu}i\not\!\partial\nu+L_{(W,Z,\psi)}
\label{fermion}\end{equation}
where $L_{(W,Z,\psi)}$ contains the interactions of the fermions with $W$
and $Z$. For example one gets
$${g\over\sqrt{2}}\left\{\bar{\nu}_L W_\mu^+\gamma^\mu e_L+
\bar{e}_L W_\mu^-\gamma^\mu \nu_L \right\}$$
which precisely describes the weak interaction mediated by the $W^\pm$ as
described above. One sees that its coupling constant is ${g/\sqrt{2}}$.
Hence in the low energy limit one recovers the four fermions interaction
with the Fermi coupling constant:
$$\left({G_F\over\sqrt{2}}\right)={g^2\over 8M_W^2}$$
Moreover one identifies the electric charge in equation~(\ref{fermion}) as:
$$e=\sin\theta_W \cos\theta_W\sqrt{g^2+g'^2}$$
so that all the parameters of the theory are related to experimental
quantities. The computation in equation~(\ref{fermion}) also produces the
couplings of the fermions to the $Z$ field, which are slightly more
complicated and we shall not write them.

The Higgs mechanism also gives masses to the fermions. It is only necessary
to introduce a Yukawa coupling between the Higgs and the fermions in the
form:
$$L_{HF}=-c\left[ \bar{\psi}_L \phi \psi_R+\bar{\psi}_R \phi^* \psi_L\right]$$
in which $c$ is some coupling constant and we recall that $\psi_L$ is
a doublet contracted with the doublet of Higgses $\phi$ while $\psi_R$ is a
scalar, hence the whole stuff is invariant under the gauge group.
When gauging away the unphysical Higgses this Lagrangian boils down to
$-c(v+h)\bar{e}e/\sqrt{2}$
hence the electron acquires a mass $m_e=cv/\sqrt{2}$ while the neutrino
remains massless. One can always adjust the coupling $c$ so as to
obtain the correct mass $m_e$ of the electron, and we see that the
coupling of the Higgs to such a particle will always be proportional to
its mass (in particular negligible for light particles). One can give
arbitrary masses to the two components of the doublet by also considering
the charge conjugate of the doublet $\psi_L$. This is the mechanism that
is used to give masses to quarks. When Majorana neutrinos are considered
the procedure becomes extremely messy, and a large number of more or less
arbitrary constants enter the game.

At this point near all the ingredients of the Standard model Lagrangian
have been introduced. What is missing is the gauge fixing term and
the associated Faddeev--Popov Lagrangian. One then faces a model in
which consistent radiative corrections can be computed. As a matter
of fact the model is in very good agreement with experiments, and the
question of whether these radiative (i.e., quantum) corrections
can be seen within present experimental precision is still
controversial~\cite{NOV}. Other points which are still under active
consideration are the question of masslessness of neutrinos (are they
Weyl or Majorana neutrinos), the quest for the top quark which is
necessary to avoid anomalies, hence keep a consistent quantum theory,
and for the Higgs which still eludes experimental evidence. From
the theoretical viewpoint a more embrassing and symmetrical theory
of strong, weak and electromagnetic interactions is still lacking.

{\noindent\bf Acknowledgements:}
We thank O. Babelon, B. Machet, C. Viallet for many discussions on the
topics of these lectures and particularly M. Capdequi Peyranere for a
critical reading of the manuscript.

%\bibliography{field}

\end{document}